\documentclass[aps,superscriptaddress,twocolumn]{revtex4}

\usepackage{graphicx,graphics,epsfig}   % Include figure files
\usepackage{dcolumn}    % Align table columns on decimal point
\usepackage{bm}         % bold math
\usepackage{amsmath}    % need for subequations
\usepackage{verbatim}   % useful for program listings
\usepackage{color}      % use if color is used in text
\usepackage{times,natbib}
\usepackage{amsmath,amsfonts,amssymb,graphics,graphics,color,times}
\usepackage{bm}
\usepackage{epstopdf}

\usepackage{latexsym}
\usepackage{amsmath}
\usepackage{amssymb}
\usepackage{amsfonts}
\usepackage{amsthm}
\usepackage{mathrsfs}
\usepackage{color,verbatim,graphics}
\DeclareMathAlphabet{\mathrsfs}{U}{rsfs}{m}{n}
\DeclareMathAlphabet{\mathpzc}{OT1}{pzc}{m}{it}
\DeclareMathAlphabet{\matheus}{U}{eus}{m}{n}
\DeclareMathAlphabet{\mathbbold}{U}{bbold}{m}{n}

\def\r1{\textbf{r}}

\def\deg{\circ}

\newcommand{\ba}{\begin{eqnarray}}
\newcommand{\ea}{\end{eqnarray}}
\newcommand{\ban}{\begin{eqnarray*}}
\newcommand{\ean}{\end{eqnarray*}}
\newcommand{\be}{\begin{equation}}
\newcommand{\ee}{\end{equation}}

\newcommand{\ket}[1]{\mathinner{|#1\rangle}}

\begin{document}

\title{Efficient deterministic giant photon phase shift from a single charged quantum dot }

\author{P. Androvitsaneas}
\affiliation{Quantum Engineering Technology Labs, H. H. Wills Physics Laboratory and
Department of Electrical \& Electronic Engineering, University of Bristol, BS8 1FD, UK}

\author{A. B. Young}
%\altaffiliation{These authors contributed equally to this work}
\affiliation{Quantum Engineering Technology Labs, H. H. Wills Physics Laboratory and
Department of Electrical \& Electronic Engineering, University of Bristol, BS8 1FD, UK}

\author{J.M. Lennon}
\affiliation{Quantum Engineering Technology Labs, H. H. Wills Physics Laboratory and
Department of Electrical \& Electronic Engineering, University of Bristol, BS8 1FD, UK}

\author{C. Schneider}
\affiliation{Technische Physik, Physikalisches Institut and Wilhelm Conrad R\"ontgen-Center for Complex Material Systems, Universit\"at W\"urzburg, Am Hubland, 97474 W\"urzburg, Germany}

\author{S. Maier}
\affiliation{Technische Physik, Physikalisches Institut and Wilhelm Conrad R\"ontgen-Center for Complex Material Systems, Universit\"at W\"urzburg, Am Hubland, 97474 W\"urzburg, Germany}

\author{J.J. Hinchliff}
\affiliation{Quantum Engineering Technology Labs, H. H. Wills Physics Laboratory and
Department of Electrical \& Electronic Engineering, University of Bristol, BS8 1FD, UK}
\affiliation{Quantum Engineering Centre for Doctoral Training, H. H. Wills Physics Laboratory and Department of Electrical and Electronic Engineering, University of Bristol, Tyndall Avenue, BS8 1FD, United Kingdom
}

\author{G.S. Atkinson}
\affiliation{Quantum Engineering Technology Labs, H. H. Wills Physics Laboratory and
Department of Electrical \& Electronic Engineering, University of Bristol, BS8 1FD, UK}

\author{E. Harbord}
\affiliation{Quantum Engineering Technology Labs, H. H. Wills Physics Laboratory and
Department of Electrical \& Electronic Engineering, University of Bristol, BS8 1FD, UK}

\author{M. Kamp}\affiliation{Technische Physik, Physikalisches Institut and Wilhelm Conrad R\"ontgen-Center for Complex Material Systems, Universit\"at W\"urzburg, Am Hubland, 97474 W\"urzburg, Germany}

\author{S. H\"ofling}\affiliation{Technische Physik, Physikalisches Institut and Wilhelm Conrad R\"ontgen-Center for Complex Material Systems, Universit\"at W\"urzburg, Am Hubland, 97474 W\"urzburg, Germany}
\affiliation{SUPA, School of Physics and Astronomy, University of St Andrews, St Andrews, KY16 9SS, United Kingdom}
%
%\author{S. Knauer}
%\affiliation{Centre for Quantum Photonics, H.H. Wills Physics Laboratory, University of Bristol,
%Tyndall Avenue, Bristol, BS8 1TL, United Kingdom}
%
%\affiliation{Department of Electrical and Electronic Engineering, University of Bristol,
%Merchant Venturers Building, Woodland Road, Bristol, BS8 1UB, UK}\author{E. Harbord}
%\affiliation{Centre for Quantum Photonics, H.H. Wills Physics Laboratory, University of Bristol,
%Tyndall Avenue, Bristol, BS8 1TL, United Kingdom}

%\author{C. Y. Hu}
%\affiliation{Department of
%Electrical and Electronic Engineering, University of Bristol,
%Merchant Venturers Building, Woodland Road, Bristol, BS8 1UB, UK}

\author{J. G. Rarity}
\affiliation{Quantum Engineering Technology Labs, H. H. Wills Physics Laboratory and
Department of Electrical \& Electronic Engineering, University of Bristol, BS8 1FD, UK}

\author{R. Oulton}
\affiliation{Quantum Engineering Technology Labs, H. H. Wills Physics Laboratory and
Department of Electrical \& Electronic Engineering, University of Bristol, BS8 1FD, UK}

\begin{abstract}
Solid-state quantum emitters have long been recognised as the ideal platform to realize integrated quantum photonic technologies. We use a self-assembled negatively charged QD in a low Q-factor photonic micropillar to demonstrate for the first time a key figure of merit for deterministic switching and spin-photon entanglement: a shift in phase of an input single photon of $>90^{o}$ with values of up to $2\pi/3$ ($120^{o}$) demonstrated. This $>\pi/2$ ($90^{o}$) measured value represents an important threshold: above this value input photons interact with the emitter deterministically. A deterministic photon-emitter interaction is the only viable scalable means to achieve several vital functionalities not possible in linear optics such as quantum switches and entanglement gates. Our experimentally determined value is limited by mode mismatch between the input laser and the cavity, QD spectral fluctuations and spin relaxation. We determine that up to $80\%$ of the collected photons have interacted with the QD and undergone a phase shift of $\pi$.
\end{abstract}

\maketitle
 The dramatic progress made in quantum dots (QD) has led to single photon sources with record efficiency and indistinguishability~\cite{Ding:2016fk,SomaschiN.:2016uq,Unsleber:16,Bennette1501256}.However, QDs are not just exploited as sources; by maximising the interaction of the QD with light, one may use the QD transition to deterministically ``switch" the phase, $\phi$, of a single photon by up to $\pi$. Here we present the first solid-state implementation that achieves this key figure of merit. Using a low quality factor (Q-factor) micropillar in the ``bad-cavity"  limit we measure a phase shift of at least $2\pi/3$ ($120^{\deg}$). Accounting for background (20\%), this corresponds to a $\pi$ phase shift of all the photons reflected from the QD-cavity system. By combining this with the selection rules for QD spin transitions, one may unlock the ability to perform efficient, high fidelity quantum entanglement operations~\cite{tu-prl-75-4710,PhysRevB.78.085307,waks:153601,PhysRevLett.104.160503,PhysRevLett.92.127902}.% Here we present the first solid-state implementation that achieves a key figure of merit towards deterministic switching and spin-photon entanglement: a shift in phase of an input single photon with values of up to $2\pi/3$ ($120^{\deg}$) is demonstrated.

There are two overarching requirements for designing a practical quantum photonic switch: firstly the passive photonic structure must possess well-defined and input and output modes facilitating efficient optical coupling. Secondly the quantum emitter should show perfect interaction (i.e. $\pi$ phase shift for every interacting photon) and minimal photon scattering into leaky modes ($\gamma$) rather than the input/output mode ($\Gamma$), i.e. a high $\beta$-factor ($\beta=\frac{\Gamma}{\Gamma+\gamma}$). Such conditions have been satisfied using atom-cavity systems~\cite{Reiserer:2013ly,Reiserer2014,Tiecke2014,PhysRevLett.111.193601}, but this has so far remained elusive for solid state quantum emitters, a more natural platform for integrable/scalable devices. In this manuscript we present a novel approach using a low Q-factor micropillar cavity often termed the ``bad cavity"  limit~\cite{Carmicheal_book}. By doing this  we ensure a well-defined, efficient input and output mode~\cite{Maier:2014ve}, in contrast to more traditional approaches using high Q-factor cavities~\cite{nat-432-7014, Yoshie:2004uq} where limits in current fabrication tolerances lead to scattering into parasitic modes, limiting the phase shift to the order of  $\sim\pi/10$~\cite{IlyaFushman05092008,Young:2011uq,Arnold:2015bh}. Recent demonstrations have exploited a similar small photon phase shift to probabilistically herald a $\pi$ phase switch on the QD spin, upon the successful detection of a rotated photon~\cite{Sun2016}. Importantly, in that case one needed to post-select on rotated photons, which occurs with a low probability. We now demonstrate the next step, showing instead a phase switch imparted by the QD onto the photons in a deterministic manner (i.e. {\it all} photons that have interacted with the QD undergo a phase shift of $\pi$). A deterministic interaction is essential in order to build large chains of entangled photons for cluster state quantum computation~\cite{Lindner:2009zr,PhysRevA.78.032318}.

A both bright and deterministic device is made possible by the unique properties of the low Q-factor micropillar design ($Q\sim290$) which inherently has negligible passive losses. Further, we have already demonstrated~\cite{Androvitsaneas:2016kx}  that these low Q-factor cavities also have a high $\beta$-factor ($\beta\sim0.65$ with up to $\beta>0.9$ possible), and therefore these structures currently represent the state of the art in terms of a practical solid-state quantum photonic switch. Our previous work ~\cite{Androvitsaneas:2016kx} has shown phase shifts $\sim6^{\deg}$, whilst we calculate that this QD-micropillar should give a $\pi$ phase shift for resonantly reflected photons (see supplement). The low phase shift actually measured in Ref.~\cite{Androvitsaneas:2016kx} is largely due to spectral ``jitter" which leads to a shift ($\delta\omega$) in the spectral position of the QD over a timescale shorter than the laser scan time~\cite{Kuhlmann:2015kx}. In those experiments with one second integration times we probe a range of QD laser detunings ($\delta\omega$), which gives rise to a much lower average photon phase shift.

In this work we overcome the spectral jitter by measuring the photon phase shift in $100\mu$s intervals (faster than the spectral jitter time). By using a specially designed two-channel heralding technique, we explore time windows when the QD remains close to resonance with the laser, and measure the phase shift of light reflected from the QD-micropillar system. This is possible due to the high brightness of the system (large $\beta$-factor, and low passive cavity losses), where we can on average detect on the order of 100 photons in 100$\mu$s, even in the weak excitation limit. We consistently evaluate phase shift values $>\pi/2$ when the QD is close to resonance ($|\delta\omega|<\Gamma/2$) where $\Gamma$ is the linewidth of the transition. We also demonstrate that in the rare cases where the laser remains precisely on-resonance ($|\delta\omega|<\Gamma/20$), we measure phase shifts of up to $8\pi/10\pm\pi/10$. Note that resonant phase shift values $\pi/2<\phi<\pi$ should not occur (see supplementary), and so measurement of $\phi>\pi/2$ reveals that the deterministic regime ($\phi=\pi$) has already been reached, but that the observable phase shift is limited/obscured by background non-interating photons (20\% of the collected signal).

%In this work we demonstrate that measuring the phase shift in $100\mu$s intervals (faster than the spectral jitter time), provides a more realistic measure of the phase response. By using a specially designed two-channel heralding technique, we explore time windows when the QD remains close to resonance with the laser, and measure the phase shift of light reflected from the QD-micropillar system. We consistently evaluate phase shift values $>\pi/2$ when the QD is close to resonance ($|\delta\omega|<\Gamma/2$) where $\Gamma$ is the linewidth of the transition. We also demonstrate that in the rare cases where we remain precisely on-resonance ($|\delta\omega|<\Gamma/20$), we measure phase shifts of up to $7\pi/10\pm\pi/10$. We therefore conclude that we reach the $>\pi/2$ threshold, confirming that $\beta>0.5$, and that only external mode-matching limits our measured phase shift.

\begin{figure}
\includegraphics[width=0.48\textwidth]{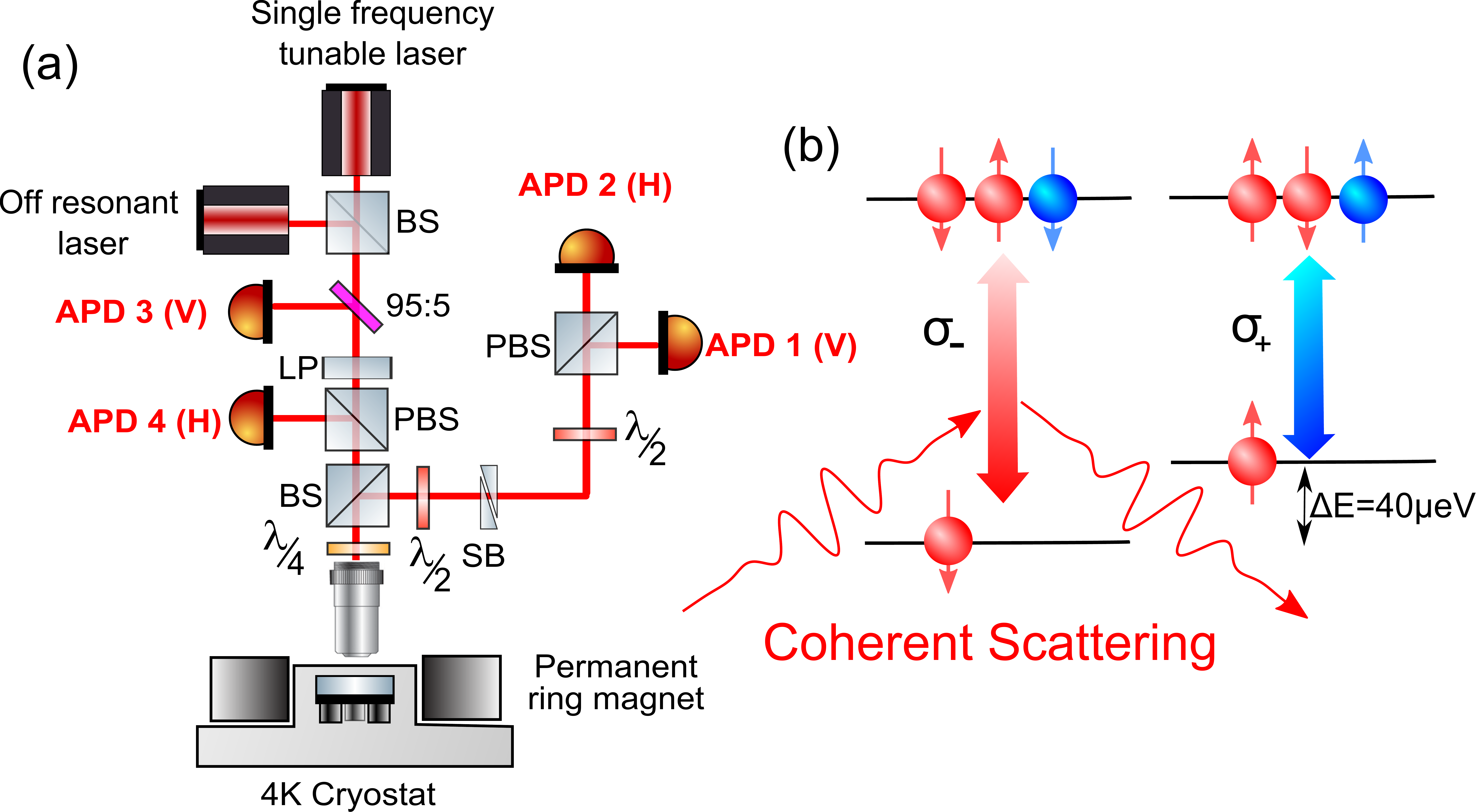}
\caption{ ({\bf a}). Experimental setup used to time resolve the QD induced phase shift. A $50:50$ beamsplitter (BS) splits the reflected signal into two channels. One which sends photons to APD-1 and APD-2 that are used as heralding detectors, with the independently correlated phase shift measured via APD-3 and APD-4. ({\bf b}) Energy level diagram of a negatively charged QD in a Faraday magnetic field. Measurements are performed on light that is resonantly scattered from the spin down transition.}\label{fig:1n}
\end{figure}

The experimental setup is shown in Fig.\ref{fig:1n}.(a), vertically linearly polarized coherent light $\ket{V}$ from a single frequency laser is input to the micropillar, and spectrally tuned to the spin $\ket{\downarrow}$ transition (see Fig.\ref{fig:1n}.(b)). The reflected signal is split into two with a non-polarizing beamsplitter (BS1): one arm is sent to APDs 3 (vertical-$V$) and 4 (horizontal-$H$); the other arm is sent to independent detectors that also measure in the linearly polarised basis APD-1 ($V$), APD-2 ($H$) (see methods). A  lower bound value for the phase shift ($cos(\phi_{LB})$) that does not take into account ellipticity (see Supplementary for full details) and underestimates the phase can be calculated via:

%The reflected signal is split into two with a non-polarizing beamsplitter (BS1): one arm is sent to APDs 3 (vertical-$V$) and 4 (horizontal-$H$); the other arm is sent to independent detectors that also measure in the linearly polarised basis APD-1 ($V$), APD-2 ($H$). A  lower bound value for the phase shift ($cos(\phi_{LB})$) that does not take into account ellipticity (see Supplementary for full details) can be calculated via:
\begin{equation}
\label{phaseLB}
 cos(\phi_{LB})=\frac{V-H}{V+H}
\end{equation}
\begin{figure}
\centering
\includegraphics[width=0.49\textwidth]{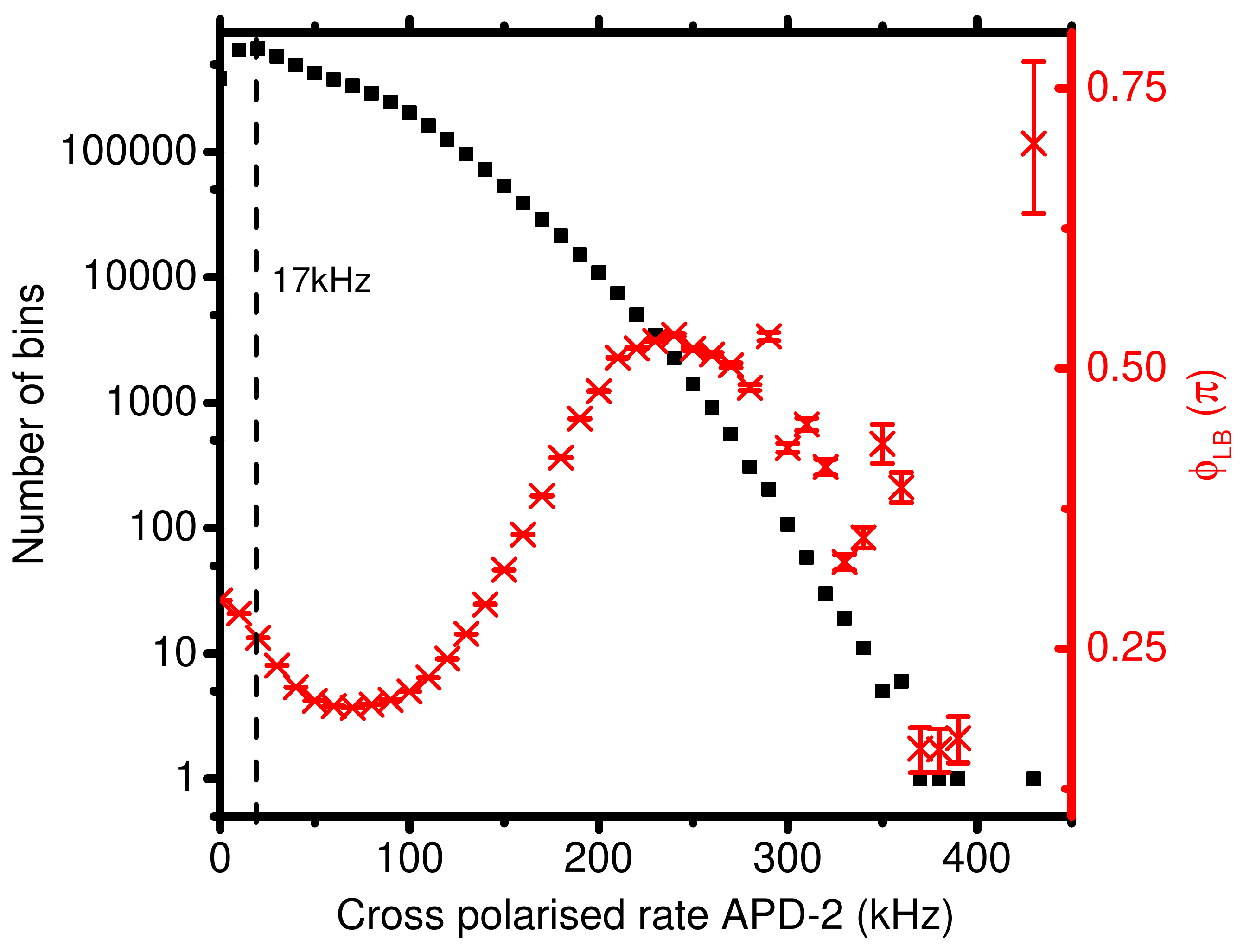}
\caption{Frequency plot of the count rates in APD2 (cross polarised) as measured in 100 $\mu$s bins (black curve) left axis. Red curve: displays the corresponding lower bound phase shift (right axis) measured by Eq.1 using APD-4 (H polarised) and APD-3 (V polarised)}\label{fig:1da}
\end{figure}
Cross-polarized counts are only observed when the laser is on-resonance with the QD and undergoes a Faraday rotation with a non-zero projection onto H. In non-cavity experiments this is the channel where cross-polarised resonant scattering is detected~\cite{Kuhlmann:2015kx}. Initially we monitor the number of counts in the cross-polarised channel via APD-2. Fig.\ref{fig:1da} shows a histogram for the distribution of count-rates in APD-2 for $100\mu$s bin widths (note the logarithmic scale). The vast majority of time bins contain low count-rates of around 17kHz. This corresponds to the QD transition shifted away from resonance with the input laser due to spectral jitter. Nevertheless there are still instances where the count-rate can become significantly higher. This increase represents light resonantly scattering from the QD transition into the cross-polarised channel.

It is particularly illuminating to correlate these count-rates with the phase measured according to Eq.~\ref{phaseLB} using APD-3 (V) and APD-4 (H). Using these detectors in a separate arm allows us to obtain a measure of the phase shift that is correlated to APD-2, but that does not unfairly bias the data we use to measure the phase shift. We expect that when the cross-polarized count-rates in the heralding arm are high, the phase shift of input photons will also be significant. In Fig.\ref{fig:1da} we can clearly see that as the number of cross-polarised counts increases so does the measured phase shift, as expected. This reaches a peak value of $\phi_{LB}=0.530\pi\pm0.001\pi$ at APD-2 count-rate 240kHz. The measured phase shift is limited by the heralding accuracy. At 240kHz we detect 24 photons per bin, corresponding to a significant Poissonian noise on the {\it x-axis} value in Fig.\ref{fig:1da}.

One can gain more information by implementing a double heralding technique. Efficient coherent scattering from the QD will result in light rotating from $\ket{V}\rightarrow\ket{H}$, causing a significant intensity reduction in the directly reflected channel (APD-1), i.e. the co-polarized counts should decrease. This should also occur just as the cross-polarized counts increase (APD-2). This anti-correlation gives a more accurate measure of when the QD is on-resonance, as it reduces the influence of uncorrelated background counts between the two channels. Note that it is not sufficient to use only APD-1 and APD-2 to infer the phase shift as even for small phase shifts there is a small statistical probability that APD-2 counts more photons than APD-1 thus we could {\it trivially postselect} a small number of $\pi$ phase shift instances. In contrast the double heralding technique here does not work as a post selection: APD-1 and APD-2 herald only that the QD is on resonance with the laser, and acts as the trigger at which point we collect all the photons in APD-3 and APD-4 (uncorrelated with APDs- 1 and 2) to measure the phase.

\begin{figure}
\centering
\includegraphics[width=0.49\textwidth]{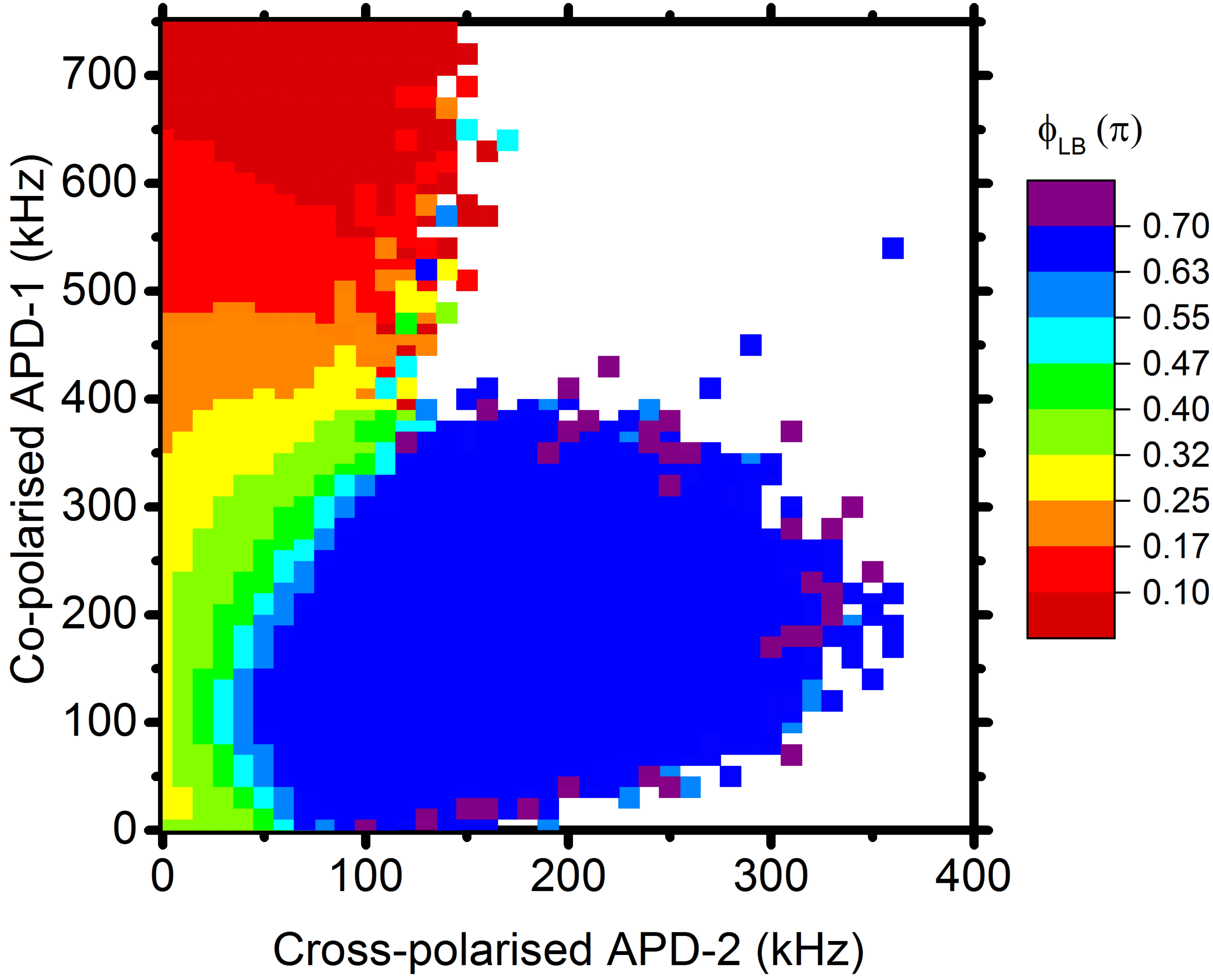}
\caption{A plot of the number of the lower bound phase shift for specific count-rates in detectors APD-1 and APD-2 in 100$\mu$s time bins. The colour map now represents the QD induced phase shift measured via detectors APD-3, and APD-4.}\label{fig:2da}
\end{figure}

To explore this Fig.~\ref{fig:2da} shows a 2D histogram revealing the correlation between the lower bound value of the phase shift ($cos(\phi_{LB})=\frac{APD 3 - APD4}{APD 3 +APD4}$) and the count-rates in the co- and cross-channels of the herald arm (APD-1,2), again, for $100\mu s$ time bins. A clear pattern emerges. For time bins where the co-polarised (APD-1) counts are high, and the cross-polarised counts (APD-2) are low (approx APD-1 $>$400kHz, APD-2 $<$ 50kHz), we observe a small phase shift $\phi_{LB}<0.1\pi$. This corresponds to the off-resonance behavior. However when the cross-polarised counts are high and the co-polarised counts are low one observes a drastic shift, represented by a large ``hot-spot" in the data (approximately, APD-2 $>$ 100kHz, APD-1 $<$ 400kHz). Here, the phase shift $\phi_{LB}$ is consistently above $0.63\pi$. This scattering occurs when the QD and laser are close to resonance. Note that it might first appear surprising that data from APDs-3,4 give a $>\pi/2$ phase shift for coordinates where APD-1$>$APD-2 (one would expect a polarisation reversal). However, this occurs due to experimental limitations. APDs- 3,4 have much lower background counts than APDs-1,2. This reduced background leads to a more accurate measure of the phase shift (see Supplementary for details).

Further, there are clearly ``pockets" in the data where the phase shift is much higher than $0.63\pi$. In particular one observes the highest phase shifts for very low $\frac{V}{H}$ ratios in the herald arm. This implies that there are specific time windows that allow the highest phase shifts to be measured. We calculate that to measure a $>0.95\pi$ phase shift, the laser should be within $\pm15neV$ of the peak of the $0.7\mu$eV linewidth transition. We calculate that this should occur $\sim0.1\%$ of the time (assuming inhomogeneous broadening of a 0.7$\mu$eV QD line to 5$\mu$ev). We also require that enough photon counts can be gathered in the time window in question to overcome Poissonian statistics, and that the $T_1$ spin lifetime of the transition is longer than the time bin used. By varying the time bin width we determine a spin $T_1$ time of 250$\mu$s and a spectral jitter time of $1.5ms$ (see Supplementary).  

Finding the instances where near exact resonance conditions are fulfilled will allow the highest phase shift possible to be measured. We can reliably measure the phase shift inside the ``hot-spot" of  $\phi_{LB}=0.687\pi\pm0.001\pi$ ($123.7^{\deg}\pm0.2^{\deg}$). The highest phase shift measured in the 100$\mu$s bins in Fig.~\ref{fig:2da} corresponds to $\phi_{LB}=0.84\pi\pm0.08\pi$ ($150^{\deg}\pm14^{\deg}$). The limit in the phase shift measured is due to ellipticity induced on the reflected signal, which occurs when $\beta<1$, along with residual background photons. In order to correct for this one needs to measure in both the linear and circular polarised basis, however practically mode-matching limitations prevent an accurate phase measurement in this case (see Supplementary). On a practical level the fact that these large phase shifts can be observed indicates that spectral jitter can easily be overcome simply by collecting sufficient photons over a timescale where the QD is stable. By increasing the collection efficiency of the scattered photons (currently $<1\%$) then it will be possible to actively trigger when the QD is on resonance as opposed to after the data has been collected as we do here.

We conclude that using a low Q-factor, high efficiency QD microcavity system, one may achieve deterministic photon-spin interactions, by inducing a $>\pi/2$ phase shift on a narrow bandwidth single photon. Not only do we measure, to our knowledge, by far the largest ever photon phase shift from a solid state quantum emitter, we reach the crucial {\it measured} threshold of $\phi > \pi/2$ ~\cite{Hofmann:2003fk}. When on resonance we infer that $80\%$ of the reflected photons in the collection channel undergo a deterministic $\pi$ phase shift. Further we have shown that the spectral jitter in these systems can be slow ($10^5$ times slower than the exciton lifetime) and that we can trigger on time windows where the QD line is stable. This would enable the generation of continuous streams of hundreds of photons, a necessary requirement for photonic cluster states\cite{Lindner:2009zr}. The significant challenge
remaining is efficient mode matching of the reflected photons. High fidelity gate operations will require exceptional reduction in background scatter implying $> 99\%$ optical mode-matching is required. By integrating these micropillars, perhaps directly with optical fibre, one could potentially reflect and collect $10^4$ photons per $100\mu$s. This would allow one to produce a 20 photon cluster state several times per second, representing a leap forward for the number of photonic qubits available for quantum information applications.

\section*{Methods}

The system under consideration is a micropillar photonic structure. The sample has been fabricated via molecular beam epitaxy (MBE) and contains a $\lambda$-thick cavity surrounded by two distributed Bragg reflectors (DBRs). The DBR structure consists of 18.5 (5) bottom (top) AlAs/GaAs mirror pairs. A modulation doped low density In(Ga)As QD layer ($1.8$x$10^9$ cm$^{-2}$) has been grown in the middle of the cavity~\cite{Maier:2014ve}. The QDs have been grown spectrally close to the cavity mode resonance at $\sim1388.6 meV$. The asymmetric low-Q design ensures that the cavity field decays predominantly through the top mirror (with a rate $>90\%$)~\cite{6493378}. In this way a high QD emission extraction efficiency is expected for these systems. From a range of different etched micropillar diameters one of a nominal diameter  $\sim2$ $\mu$m was selected. We find that the fundamental mode of the cavity appears at $1388$ meV  cavity and has a full width half maximum of $\sim4.1$meV corresponding to a Q-factor of $Q\sim290$. The fundamental mode is well separated from any higher modes that appear typically $\gtrsim16$meV to the blue side of the fundamental mode. The cumulative rate of losses into higher order modes are calculated to be not more than 3\% of the rate of emission into the fundamental mode. The high light collection efficiency is evident from measured QD count rates of 1 MCount/sec at saturation under resonant excitation.

The setup used is shown in Fig.~\ref{fig:1n}, and is similar to that in Ref.~\cite{Androvitsaneas:2016kx}. The sample is placed in a $0.5T$ magnetic field along the optical axis (Faraday geometry) introduced by the use of a permanent ring magnet, which creates a homogeneous magnetic field perpendicular to the plane of the QD. This induces an energy splitting of $\sim40\mu eV$ between the spin up ($\ket{\uparrow}$) and down($\ket{\downarrow}$) ground state (Fig 1b). Vertically linearly polarized coherent light ($\ket{V}=\frac{1}{\sqrt{2}}(\ket{R}-i\ket{L})$) from a single frequency laser of linewidth $10MHz$ (over $1ms$) is focused on the micropillar with an $NA= 0.7$ objective. It is then spectrally tuned to the spin $\ket{\downarrow}$ transition, which interacts with $\ket{L}$ light. This imposes a selective phase shift on only one circular component $\ket{L}$ of the linearly polalised photon ($\ket{V}$), implementing the transform $\ket{V}\rightarrow 1/\sqrt{2}(\ket{R}-ie^{i\phi}\ket{L})$. For the case when $\beta>0.5$, $\phi = \pi$, which leads to a ``giant Faraday rotation" ($\phi_{FR}=\phi/2=\pi/2$), switching the polarisation state from $\ket{V}$, to $\ket{H}$~\cite{PhysRevB.78.085307}. Spin-photon entanglement naturally ensues if we can prepare the spin into an equal superposition state: $\ket{\psi}=\frac{1}{\sqrt{2}}(\ket{\uparrow}+\ket{\downarrow})$.

\bibliography{bib13deskpetros}

\begin{thebibliography}{27}
\expandafter\ifx\csname natexlab\endcsname\relax\def\natexlab#1{#1}\fi
\expandafter\ifx\csname bibnamefont\endcsname\relax
  \def\bibnamefont#1{#1}\fi
\expandafter\ifx\csname bibfnamefont\endcsname\relax
  \def\bibfnamefont#1{#1}\fi
\expandafter\ifx\csname citenamefont\endcsname\relax
  \def\citenamefont#1{#1}\fi
\expandafter\ifx\csname url\endcsname\relax
  \def\url#1{\texttt{#1}}\fi
\expandafter\ifx\csname urlprefix\endcsname\relax\def\urlprefix{URL }\fi
\providecommand{\bibinfo}[2]{#2}
\providecommand{\eprint}[2][]{\url{#2}}

\bibitem[{\citenamefont{Ding et~al.}(2016)\citenamefont{Ding, He, Duan,
  Gregersen, Chen, Unsleber, Maier, Schneider, Kamp, H{\"o}fling
  et~al.}}]{Ding:2016fk}
\bibinfo{author}{\bibfnamefont{X.}~\bibnamefont{Ding}},
  \bibinfo{author}{\bibfnamefont{Y.}~\bibnamefont{He}},
  \bibinfo{author}{\bibfnamefont{Z.~C.} \bibnamefont{Duan}},
  \bibinfo{author}{\bibfnamefont{N.}~\bibnamefont{Gregersen}},
  \bibinfo{author}{\bibfnamefont{M.~C.} \bibnamefont{Chen}},
  \bibinfo{author}{\bibfnamefont{S.}~\bibnamefont{Unsleber}},
  \bibinfo{author}{\bibfnamefont{S.}~\bibnamefont{Maier}},
  \bibinfo{author}{\bibfnamefont{C.}~\bibnamefont{Schneider}},
  \bibinfo{author}{\bibfnamefont{M.}~\bibnamefont{Kamp}},
  \bibinfo{author}{\bibfnamefont{S.}~\bibnamefont{H{\"o}fling}},
  \bibnamefont{et~al.}, \bibinfo{journal}{Physical Review Letters}
  \textbf{\bibinfo{volume}{116}}, \bibinfo{pages}{020401}
  (\bibinfo{year}{2016}),
  \urlprefix\url{http://link.aps.org/doi/10.1103/PhysRevLett.116.020401}.

\bibitem[{\citenamefont{SomaschiN. et~al.}(2016)\citenamefont{SomaschiN.,
  GieszV., SantisL., C., P., HorneckerG., L., GrangeT., Ant{\'o}nC., DemoryJ.
  et~al.}}]{SomaschiN.:2016uq}
\bibinfo{author}{\bibnamefont{SomaschiN.}},
  \bibinfo{author}{\bibnamefont{GieszV.}},
  \bibinfo{author}{\bibfnamefont{D.}~\bibnamefont{SantisL.}},
  \bibinfo{author}{\bibfnamefont{L.}~\bibnamefont{C.}},
  \bibinfo{author}{\bibfnamefont{A.}~\bibnamefont{P.}},
  \bibinfo{author}{\bibnamefont{HorneckerG.}},
  \bibinfo{author}{\bibfnamefont{P.}~\bibnamefont{L.}},
  \bibinfo{author}{\bibnamefont{GrangeT.}},
  \bibinfo{author}{\bibnamefont{Ant{\'o}nC.}},
  \bibinfo{author}{\bibnamefont{DemoryJ.}}, \bibnamefont{et~al.},
  \bibinfo{journal}{Nat Photon} \textbf{\bibinfo{volume}{10}},
  \bibinfo{pages}{340} (\bibinfo{year}{2016}),
  \urlprefix\url{http://dx.doi.org/10.1038/nphoton.2016.23}.

\bibitem[{\citenamefont{Unsleber et~al.}(2016)\citenamefont{Unsleber, He,
  Gerhardt, Maier, Lu, Pan, Gregersen, Kamp, Schneider, and
  H\"{o}fling}}]{Unsleber:16}
\bibinfo{author}{\bibfnamefont{S.}~\bibnamefont{Unsleber}},
  \bibinfo{author}{\bibfnamefont{Y.-M.} \bibnamefont{He}},
  \bibinfo{author}{\bibfnamefont{S.}~\bibnamefont{Gerhardt}},
  \bibinfo{author}{\bibfnamefont{S.}~\bibnamefont{Maier}},
  \bibinfo{author}{\bibfnamefont{C.-Y.} \bibnamefont{Lu}},
  \bibinfo{author}{\bibfnamefont{J.-W.} \bibnamefont{Pan}},
  \bibinfo{author}{\bibfnamefont{N.}~\bibnamefont{Gregersen}},
  \bibinfo{author}{\bibfnamefont{M.}~\bibnamefont{Kamp}},
  \bibinfo{author}{\bibfnamefont{C.}~\bibnamefont{Schneider}},
  \bibnamefont{and}
  \bibinfo{author}{\bibfnamefont{S.}~\bibnamefont{H\"{o}fling}},
  \bibinfo{journal}{Opt. Express} \textbf{\bibinfo{volume}{24}},
  \bibinfo{pages}{8539} (\bibinfo{year}{2016}),
  \urlprefix\url{http://www.opticsexpress.org/abstract.cfm?URI=oe-24-8-8539}.

\bibitem[{\citenamefont{Bennett et~al.}(2016)\citenamefont{Bennett, Lee, Ellis,
  Meany, Murray, Floether, Griffths, Farrer, Ritchie, and
  Shields}}]{Bennette1501256}
\bibinfo{author}{\bibfnamefont{A.~J.} \bibnamefont{Bennett}},
  \bibinfo{author}{\bibfnamefont{J.~P.} \bibnamefont{Lee}},
  \bibinfo{author}{\bibfnamefont{D.~J.~P.} \bibnamefont{Ellis}},
  \bibinfo{author}{\bibfnamefont{T.}~\bibnamefont{Meany}},
  \bibinfo{author}{\bibfnamefont{E.}~\bibnamefont{Murray}},
  \bibinfo{author}{\bibfnamefont{F.~F.} \bibnamefont{Floether}},
  \bibinfo{author}{\bibfnamefont{J.~P.} \bibnamefont{Griffths}},
  \bibinfo{author}{\bibfnamefont{I.}~\bibnamefont{Farrer}},
  \bibinfo{author}{\bibfnamefont{D.~A.} \bibnamefont{Ritchie}},
  \bibnamefont{and} \bibinfo{author}{\bibfnamefont{A.~J.}
  \bibnamefont{Shields}}, \bibinfo{journal}{Science Advances}
  \textbf{\bibinfo{volume}{2}} (\bibinfo{year}{2016}),
  \eprint{http://advances.sciencemag.org/content/2/4/e1501256.full.pdf},
  \urlprefix\url{http://advances.sciencemag.org/content/2/4/e1501256}.

\bibitem[{\citenamefont{Turchette et~al.}(1995)\citenamefont{Turchette, Hood,
  Lange, Mabuchi, and Kimble}}]{tu-prl-75-4710}
\bibinfo{author}{\bibfnamefont{Q.~A.} \bibnamefont{Turchette}},
  \bibinfo{author}{\bibfnamefont{C.~J.} \bibnamefont{Hood}},
  \bibinfo{author}{\bibfnamefont{W.}~\bibnamefont{Lange}},
  \bibinfo{author}{\bibfnamefont{H.}~\bibnamefont{Mabuchi}}, \bibnamefont{and}
  \bibinfo{author}{\bibfnamefont{H.~J.} \bibnamefont{Kimble}},
  \bibinfo{journal}{Phys. Rev. Lett.} \textbf{\bibinfo{volume}{75}},
  \bibinfo{pages}{4710} (\bibinfo{year}{1995}).

\bibitem[{\citenamefont{Hu et~al.}(2008)\citenamefont{Hu, Young, O'Brien,
  Munro, and Rarity}}]{PhysRevB.78.085307}
\bibinfo{author}{\bibfnamefont{C.~Y.} \bibnamefont{Hu}},
  \bibinfo{author}{\bibfnamefont{A.}~\bibnamefont{Young}},
  \bibinfo{author}{\bibfnamefont{J.~L.} \bibnamefont{O'Brien}},
  \bibinfo{author}{\bibfnamefont{W.~J.} \bibnamefont{Munro}}, \bibnamefont{and}
  \bibinfo{author}{\bibfnamefont{J.~G.} \bibnamefont{Rarity}},
  \bibinfo{journal}{Phys. Rev. B} \textbf{\bibinfo{volume}{78}},
  \bibinfo{pages}{085307} (\bibinfo{year}{2008}).

\bibitem[{\citenamefont{Waks and Vuckovic}(2006)}]{waks:153601}
\bibinfo{author}{\bibfnamefont{E.}~\bibnamefont{Waks}} \bibnamefont{and}
  \bibinfo{author}{\bibfnamefont{J.}~\bibnamefont{Vuckovic}},
  \bibinfo{journal}{Phys. Rev. Lett.} \textbf{\bibinfo{volume}{96}},
  \bibinfo{eid}{153601} (pages~\bibinfo{numpages}{4}) (\bibinfo{year}{2006}),
  \urlprefix\url{http://link.aps.org/abstract/PRL/v96/e153601}.

\bibitem[{\citenamefont{Bonato et~al.}(2010)\citenamefont{Bonato, Haupt,
  Oemrawsingh, Gudat, Ding, van Exter, and
  Bouwmeester}}]{PhysRevLett.104.160503}
\bibinfo{author}{\bibfnamefont{C.}~\bibnamefont{Bonato}},
  \bibinfo{author}{\bibfnamefont{F.}~\bibnamefont{Haupt}},
  \bibinfo{author}{\bibfnamefont{S.~S.~R.} \bibnamefont{Oemrawsingh}},
  \bibinfo{author}{\bibfnamefont{J.}~\bibnamefont{Gudat}},
  \bibinfo{author}{\bibfnamefont{D.}~\bibnamefont{Ding}},
  \bibinfo{author}{\bibfnamefont{M.~P.} \bibnamefont{van Exter}},
  \bibnamefont{and}
  \bibinfo{author}{\bibfnamefont{D.}~\bibnamefont{Bouwmeester}},
  \bibinfo{journal}{Phys. Rev. Lett.} \textbf{\bibinfo{volume}{104}},
  \bibinfo{pages}{160503} (\bibinfo{year}{2010}).

\bibitem[{\citenamefont{Duan and Kimble}(2004)}]{PhysRevLett.92.127902}
\bibinfo{author}{\bibfnamefont{L.-M.} \bibnamefont{Duan}} \bibnamefont{and}
  \bibinfo{author}{\bibfnamefont{H.~J.} \bibnamefont{Kimble}},
  \bibinfo{journal}{Phys. Rev. Lett.} \textbf{\bibinfo{volume}{92}},
  \bibinfo{pages}{127902} (\bibinfo{year}{2004}).

\bibitem[{\citenamefont{Reiserer et~al.}(2013)\citenamefont{Reiserer, Ritter,
  and Rempe}}]{Reiserer:2013ly}
\bibinfo{author}{\bibfnamefont{A.}~\bibnamefont{Reiserer}},
  \bibinfo{author}{\bibfnamefont{S.}~\bibnamefont{Ritter}}, \bibnamefont{and}
  \bibinfo{author}{\bibfnamefont{G.}~\bibnamefont{Rempe}},
  \bibinfo{journal}{Science} \textbf{\bibinfo{volume}{342}},
  \bibinfo{pages}{1349} (\bibinfo{year}{2013}),
  \urlprefix\url{http://www.sciencemag.org/content/342/6164/1349}.

\bibitem[{\citenamefont{Reiserer et~al.}(2014)\citenamefont{Reiserer, Kalb,
  Rempe, and Ritter}}]{Reiserer2014}
\bibinfo{author}{\bibfnamefont{A.}~\bibnamefont{Reiserer}},
  \bibinfo{author}{\bibfnamefont{N.}~\bibnamefont{Kalb}},
  \bibinfo{author}{\bibfnamefont{G.}~\bibnamefont{Rempe}}, \bibnamefont{and}
  \bibinfo{author}{\bibfnamefont{S.}~\bibnamefont{Ritter}},
  \bibinfo{journal}{Nature} \textbf{\bibinfo{volume}{508}},
  \bibinfo{pages}{237} (\bibinfo{year}{2014}), ISSN \bibinfo{issn}{0028-0836},
  \bibinfo{note}{letter},
  \urlprefix\url{http://dx.doi.org/10.1038/nature13177}.

\bibitem[{\citenamefont{Tiecke et~al.}(2014)\citenamefont{Tiecke, Thompson,
  de~Leon, Liu, Vuletic, and Lukin}}]{Tiecke2014}
\bibinfo{author}{\bibfnamefont{T.~G.} \bibnamefont{Tiecke}},
  \bibinfo{author}{\bibfnamefont{J.~D.} \bibnamefont{Thompson}},
  \bibinfo{author}{\bibfnamefont{N.~P.} \bibnamefont{de~Leon}},
  \bibinfo{author}{\bibfnamefont{L.~R.} \bibnamefont{Liu}},
  \bibinfo{author}{\bibfnamefont{V.}~\bibnamefont{Vuletic}}, \bibnamefont{and}
  \bibinfo{author}{\bibfnamefont{M.~D.} \bibnamefont{Lukin}},
  \bibinfo{journal}{Nature} \textbf{\bibinfo{volume}{508}},
  \bibinfo{pages}{241} (\bibinfo{year}{2014}), ISSN \bibinfo{issn}{0028-0836},
  \bibinfo{note}{letter},
  \urlprefix\url{http://dx.doi.org/10.1038/nature13188}.

\bibitem[{\citenamefont{O'Shea et~al.}(2013)\citenamefont{O'Shea, Junge, Volz,
  and Rauschenbeutel}}]{PhysRevLett.111.193601}
\bibinfo{author}{\bibfnamefont{D.}~\bibnamefont{O'Shea}},
  \bibinfo{author}{\bibfnamefont{C.}~\bibnamefont{Junge}},
  \bibinfo{author}{\bibfnamefont{J.}~\bibnamefont{Volz}}, \bibnamefont{and}
  \bibinfo{author}{\bibfnamefont{A.}~\bibnamefont{Rauschenbeutel}},
  \bibinfo{journal}{Phys. Rev. Lett.} \textbf{\bibinfo{volume}{111}},
  \bibinfo{pages}{193601} (\bibinfo{year}{2013}),
  \urlprefix\url{https://link.aps.org/doi/10.1103/PhysRevLett.111.193601}.

\bibitem[{\citenamefont{Carmichael}(2008)}]{Carmicheal_book}
\bibinfo{author}{\bibfnamefont{H.~J.} \bibnamefont{Carmichael}},
  \emph{\bibinfo{title}{Statistical Methods in Quantum Optics 2}}
  (\bibinfo{publisher}{Springer-Verlag}, \bibinfo{year}{2008}).

\bibitem[{\citenamefont{Maier et~al.}(2014)\citenamefont{Maier, Gold, Forchel,
  Gregersen, M{\o}rk, H{\"o}fling, Schneider, and Kamp}}]{Maier:2014ve}
\bibinfo{author}{\bibfnamefont{S.}~\bibnamefont{Maier}},
  \bibinfo{author}{\bibfnamefont{P.}~\bibnamefont{Gold}},
  \bibinfo{author}{\bibfnamefont{A.}~\bibnamefont{Forchel}},
  \bibinfo{author}{\bibfnamefont{N.}~\bibnamefont{Gregersen}},
  \bibinfo{author}{\bibfnamefont{J.}~\bibnamefont{M{\o}rk}},
  \bibinfo{author}{\bibfnamefont{S.}~\bibnamefont{H{\"o}fling}},
  \bibinfo{author}{\bibfnamefont{C.}~\bibnamefont{Schneider}},
  \bibnamefont{and} \bibinfo{author}{\bibfnamefont{M.}~\bibnamefont{Kamp}},
  \bibinfo{journal}{Optics Express} \textbf{\bibinfo{volume}{22}},
  \bibinfo{pages}{8136} (\bibinfo{year}{2014}),
  \urlprefix\url{http://www.opticsexpress.org/abstract.cfm?URI=oe-22-7-8136}.

\bibitem[{\citenamefont{Reithmaier.~et al}(2004)}]{nat-432-7014}
\bibinfo{author}{\bibfnamefont{J.~P.} \bibnamefont{Reithmaier.~et al}},
  \bibinfo{journal}{Nature} \textbf{\bibinfo{volume}{432}},
  \bibinfo{pages}{197} (\bibinfo{year}{2004}),
  \urlprefix\url{http://dx.doi.org/10.1038/nature02969}.

\bibitem[{\citenamefont{Yoshie et~al.}(2004)\citenamefont{Yoshie, Scherer,
  Hendrickson, Khitrova, Gibbs, Rupper, Ell, Shchekin, and
  Deppe}}]{Yoshie:2004uq}
\bibinfo{author}{\bibfnamefont{T.}~\bibnamefont{Yoshie}},
  \bibinfo{author}{\bibfnamefont{A.}~\bibnamefont{Scherer}},
  \bibinfo{author}{\bibfnamefont{J.}~\bibnamefont{Hendrickson}},
  \bibinfo{author}{\bibfnamefont{G.}~\bibnamefont{Khitrova}},
  \bibinfo{author}{\bibfnamefont{H.~M.} \bibnamefont{Gibbs}},
  \bibinfo{author}{\bibfnamefont{G.}~\bibnamefont{Rupper}},
  \bibinfo{author}{\bibfnamefont{C.}~\bibnamefont{Ell}},
  \bibinfo{author}{\bibfnamefont{O.~B.} \bibnamefont{Shchekin}},
  \bibnamefont{and} \bibinfo{author}{\bibfnamefont{D.~G.} \bibnamefont{Deppe}},
  \bibinfo{journal}{Nature} \textbf{\bibinfo{volume}{432}},
  \bibinfo{pages}{200} (\bibinfo{year}{2004}),
  \urlprefix\url{http://dx.doi.org/10.1038/nature03119}.

\bibitem[{\citenamefont{Fushman et~al.}(2008)\citenamefont{Fushman, Englund,
  Faraon, Stoltz, Petroff, and Vuckovic}}]{IlyaFushman05092008}
\bibinfo{author}{\bibfnamefont{I.}~\bibnamefont{Fushman}},
  \bibinfo{author}{\bibfnamefont{D.}~\bibnamefont{Englund}},
  \bibinfo{author}{\bibfnamefont{A.}~\bibnamefont{Faraon}},
  \bibinfo{author}{\bibfnamefont{N.}~\bibnamefont{Stoltz}},
  \bibinfo{author}{\bibfnamefont{P.}~\bibnamefont{Petroff}}, \bibnamefont{and}
  \bibinfo{author}{\bibfnamefont{J.}~\bibnamefont{Vuckovic}},
  \bibinfo{journal}{Science} \textbf{\bibinfo{volume}{320}},
  \bibinfo{pages}{769} (\bibinfo{year}{2008}),
  \eprint{http://www.sciencemag.org/cgi/reprint/320/5877/769.pdf},
  \urlprefix\url{http://www.sciencemag.org/cgi/content/abstract/320/5877/769}.

\bibitem[{\citenamefont{Young et~al.}(2011)\citenamefont{Young, Oulton, Hu,
  Thijssen, Schneider, Reitzenstein, Kamp, H{\"o}fling, Worschech, Forchel
  et~al.}}]{Young:2011uq}
\bibinfo{author}{\bibfnamefont{A.~B.} \bibnamefont{Young}},
  \bibinfo{author}{\bibfnamefont{R.}~\bibnamefont{Oulton}},
  \bibinfo{author}{\bibfnamefont{C.~Y.} \bibnamefont{Hu}},
  \bibinfo{author}{\bibfnamefont{A.~C.~T.} \bibnamefont{Thijssen}},
  \bibinfo{author}{\bibfnamefont{C.}~\bibnamefont{Schneider}},
  \bibinfo{author}{\bibfnamefont{S.}~\bibnamefont{Reitzenstein}},
  \bibinfo{author}{\bibfnamefont{M.}~\bibnamefont{Kamp}},
  \bibinfo{author}{\bibfnamefont{S.}~\bibnamefont{H{\"o}fling}},
  \bibinfo{author}{\bibfnamefont{L.}~\bibnamefont{Worschech}},
  \bibinfo{author}{\bibfnamefont{A.}~\bibnamefont{Forchel}},
  \bibnamefont{et~al.}, \bibinfo{journal}{Physical Review A}
  \textbf{\bibinfo{volume}{84}}, \bibinfo{pages}{011803}
  (\bibinfo{year}{2011}),
  \urlprefix\url{http://link.aps.org/doi/10.1103/PhysRevA.84.011803}.

\bibitem[{\citenamefont{Arnold et~al.}(2015)\citenamefont{Arnold, Demory, Loo,
  Lema{\^\i}tre, Sagnes, Glazov, Krebs, Voisin, Senellart, and
  Lanco}}]{Arnold:2015bh}
\bibinfo{author}{\bibfnamefont{C.}~\bibnamefont{Arnold}},
  \bibinfo{author}{\bibfnamefont{J.}~\bibnamefont{Demory}},
  \bibinfo{author}{\bibfnamefont{V.}~\bibnamefont{Loo}},
  \bibinfo{author}{\bibfnamefont{A.}~\bibnamefont{Lema{\^\i}tre}},
  \bibinfo{author}{\bibfnamefont{I.}~\bibnamefont{Sagnes}},
  \bibinfo{author}{\bibfnamefont{M.}~\bibnamefont{Glazov}},
  \bibinfo{author}{\bibfnamefont{O.}~\bibnamefont{Krebs}},
  \bibinfo{author}{\bibfnamefont{P.}~\bibnamefont{Voisin}},
  \bibinfo{author}{\bibfnamefont{P.}~\bibnamefont{Senellart}},
  \bibnamefont{and} \bibinfo{author}{\bibfnamefont{L.}~\bibnamefont{Lanco}},
  \bibinfo{journal}{Nat Commun} \textbf{\bibinfo{volume}{6}}
  (\bibinfo{year}{2015}), \urlprefix\url{http://dx.doi.org/10.1038/ncomms7236}.

\bibitem[{\citenamefont{Sun et~al.}(2016)\citenamefont{Sun, Kim, Solomon, and
  Waks}}]{Sun2016}
\bibinfo{author}{\bibfnamefont{S.}~\bibnamefont{Sun}},
  \bibinfo{author}{\bibfnamefont{H.}~\bibnamefont{Kim}},
  \bibinfo{author}{\bibfnamefont{G.~S.} \bibnamefont{Solomon}},
  \bibnamefont{and} \bibinfo{author}{\bibfnamefont{E.}~\bibnamefont{Waks}},
  \bibinfo{journal}{Nat Nano} \textbf{\bibinfo{volume}{11}},
  \bibinfo{pages}{539} (\bibinfo{year}{2016}), ISSN \bibinfo{issn}{1748-3387},
  \bibinfo{note}{article},
  \urlprefix\url{http://dx.doi.org/10.1038/nnano.2015.334}.

\bibitem[{\citenamefont{Lindner and Rudolph}(2009)}]{Lindner:2009zr}
\bibinfo{author}{\bibfnamefont{N.~H.} \bibnamefont{Lindner}} \bibnamefont{and}
  \bibinfo{author}{\bibfnamefont{T.}~\bibnamefont{Rudolph}},
  \bibinfo{journal}{Phys. Rev. Lett.} \textbf{\bibinfo{volume}{103}}
  (\bibinfo{year}{2009}),
  \urlprefix\url{http://link.aps.org/doi/10.1103/PhysRevLett.103.113602}.

\bibitem[{\citenamefont{Stephens et~al.}(2008)\citenamefont{Stephens, Evans,
  Devitt, Greentree, Fowler, Munro, O'Brien, Nemoto, and
  Hollenberg}}]{PhysRevA.78.032318}
\bibinfo{author}{\bibfnamefont{A.~M.} \bibnamefont{Stephens}},
  \bibinfo{author}{\bibfnamefont{Z.~W.~E.} \bibnamefont{Evans}},
  \bibinfo{author}{\bibfnamefont{S.~J.} \bibnamefont{Devitt}},
  \bibinfo{author}{\bibfnamefont{A.~D.} \bibnamefont{Greentree}},
  \bibinfo{author}{\bibfnamefont{A.~G.} \bibnamefont{Fowler}},
  \bibinfo{author}{\bibfnamefont{W.~J.} \bibnamefont{Munro}},
  \bibinfo{author}{\bibfnamefont{J.~L.} \bibnamefont{O'Brien}},
  \bibinfo{author}{\bibfnamefont{K.}~\bibnamefont{Nemoto}}, \bibnamefont{and}
  \bibinfo{author}{\bibfnamefont{L.~C.~L.} \bibnamefont{Hollenberg}},
  \bibinfo{journal}{Phys. Rev. A} \textbf{\bibinfo{volume}{78}},
  \bibinfo{pages}{032318} (\bibinfo{year}{2008}).

\bibitem[{\citenamefont{Androvitsaneas
  et~al.}(2016)\citenamefont{Androvitsaneas, Young, Schneider, Maier, Kamp,
  H{\"o}fling, Knauer, Harbord, Hu, Rarity et~al.}}]{Androvitsaneas:2016kx}
\bibinfo{author}{\bibfnamefont{P.}~\bibnamefont{Androvitsaneas}},
  \bibinfo{author}{\bibfnamefont{A.~B.} \bibnamefont{Young}},
  \bibinfo{author}{\bibfnamefont{C.}~\bibnamefont{Schneider}},
  \bibinfo{author}{\bibfnamefont{S.}~\bibnamefont{Maier}},
  \bibinfo{author}{\bibfnamefont{M.}~\bibnamefont{Kamp}},
  \bibinfo{author}{\bibfnamefont{S.}~\bibnamefont{H{\"o}fling}},
  \bibinfo{author}{\bibfnamefont{S.}~\bibnamefont{Knauer}},
  \bibinfo{author}{\bibfnamefont{E.}~\bibnamefont{Harbord}},
  \bibinfo{author}{\bibfnamefont{C.~Y.} \bibnamefont{Hu}},
  \bibinfo{author}{\bibfnamefont{J.~G.} \bibnamefont{Rarity}},
  \bibnamefont{et~al.}, \bibinfo{journal}{Physical Review B}
  \textbf{\bibinfo{volume}{93}}, \bibinfo{pages}{241409}
  (\bibinfo{year}{2016}),
  \urlprefix\url{http://link.aps.org/doi/10.1103/PhysRevB.93.241409}.

\bibitem[{\citenamefont{Kuhlmann et~al.}(2015)\citenamefont{Kuhlmann, Prechtel,
  Houel, Ludwig, Reuter, Wieck, and Warburton}}]{Kuhlmann:2015kx}
\bibinfo{author}{\bibfnamefont{A.~V.} \bibnamefont{Kuhlmann}},
  \bibinfo{author}{\bibfnamefont{J.~H.} \bibnamefont{Prechtel}},
  \bibinfo{author}{\bibfnamefont{J.}~\bibnamefont{Houel}},
  \bibinfo{author}{\bibfnamefont{A.}~\bibnamefont{Ludwig}},
  \bibinfo{author}{\bibfnamefont{D.}~\bibnamefont{Reuter}},
  \bibinfo{author}{\bibfnamefont{A.~D.} \bibnamefont{Wieck}}, \bibnamefont{and}
  \bibinfo{author}{\bibfnamefont{R.~J.} \bibnamefont{Warburton}},
  \bibinfo{journal}{Nat Commun} \textbf{\bibinfo{volume}{6}}
  (\bibinfo{year}{2015}), \urlprefix\url{http://dx.doi.org/10.1038/ncomms9204}.

\bibitem[{\citenamefont{Hofmann et~al.}(2003)\citenamefont{Hofmann, Kojima,
  Takeuchi, and Sasaki}}]{Hofmann:2003fk}
\bibinfo{author}{\bibfnamefont{H.~F.} \bibnamefont{Hofmann}},
  \bibinfo{author}{\bibfnamefont{K.}~\bibnamefont{Kojima}},
  \bibinfo{author}{\bibfnamefont{S.}~\bibnamefont{Takeuchi}}, \bibnamefont{and}
  \bibinfo{author}{\bibfnamefont{K.}~\bibnamefont{Sasaki}},
  \bibinfo{journal}{Journal of Optics B: Quantum and Semiclassical Optics}
  \textbf{\bibinfo{volume}{5}} (\bibinfo{year}{2003}),
  \urlprefix\url{http://stacks.iop.org/1464-4266/5/i=3/a=304}.

\bibitem[{\citenamefont{Gregersen et~al.}(2013)\citenamefont{Gregersen, Kaer,
  and Mørk}}]{6493378}
\bibinfo{author}{\bibfnamefont{N.}~\bibnamefont{Gregersen}},
  \bibinfo{author}{\bibfnamefont{P.}~\bibnamefont{Kaer}}, \bibnamefont{and}
  \bibinfo{author}{\bibfnamefont{J.}~\bibnamefont{Mørk}},
  \bibinfo{journal}{IEEE Journal of Selected Topics in Quantum Electronics}
  \textbf{\bibinfo{volume}{19}}, \bibinfo{pages}{1} (\bibinfo{year}{2013}),
  ISSN \bibinfo{issn}{1077-260X}.

\end{thebibliography}

\begin{acknowledgements}
The authors would like to thank X.Ai and H.F. Hofmann for helpful discussions. This work was funded by the Future Emerging Technologies (FET)-Open FP7-284743 [project Spin Photon Angular Momentum Transfer for Quantum Enabled Technologies (SPANGL4Q)] and the German Ministry of Education and research (BMBF) and Engineering and Physical Sciences Research Council (EPSRC) (EP/M024156/1, EP/N003381/1 and EP/M024458/1). J.J.H. was supported by the Bristol Quantum Engineering Centre for Doctoral Training, EPSRC grant EP/L015730/1
\end{acknowledgements}
\end{document}